\documentclass[12pt]{article}
\usepackage{amssymb}
\usepackage{epsfig}

\catcode`\@=11
\def\numberbysection{\@addtoreset{equation}{section}
        \def\theequation{\thesection.\arabic{equation}}}

\def\beq{\begin{equation}}
\def\eeq{\end{equation}}
\numberbysection
\begin{document}
\begin{titlepage}
\begin{center}
\hfill  \\
\vskip 1.in {\Large \bf Canonical transformations and minimal length} \vskip 0.5in P. Valtancoli
\\[.2in]
{\em Dipartimento di Fisica, Polo Scientifico Universit\'a di Firenze \\
and INFN, Sezione di Firenze (Italy)\\
Via G. Sansone 1, 50019 Sesto Fiorentino, Italy}
\end{center}
\vskip .5in
\begin{abstract}
We show how to modify the canonical transformations to make them compatible with non-commutative Poisson brackets.
\end{abstract}
\medskip
\end{titlepage}
\pagenumbering{arabic}
\section{Introduction}

All the various candidates of quantum gravity like string theory, loop quantum gravity and quantum geometry have one fundamental aspect in common, i.e. the introduction of a minimal observable length in a quantum system \cite{1}.

It has been observed in \cite{2}-\cite{3} that this notion can be described in a quantum mechanical language as a non-vanishing minimal uncertainty in position measurements, by substituting the Heisenberg uncertainty principle with a generalized one ( GUP ).

In this article we will analyze in detail all the implications that this choice produces at a classical level, i.e. the substitution of traditional canonical transformations and Poisson brackets with a new classical scheme which prepares the non-commutative quantization.

This analysis is useful because, as we will see in the following, the quantization schemes as the path integral are deeply based on the results of classical analytic mechanics. Therefore if we want to learn how to modify the path integral in the non-commutative direction, it is important to know its effects at a classical level.

The article is organized as follows. We start by analyzing  systems with one spatial dimension, in which case it is easy to modify the canonical transformations to make them compatible with the non-commutative Poison brackets. The general theory is then applied to the classical harmonic oscillator with a minimal length. Then we extend this procedure to the case of several spatial dimensions with the introduction of the Snyder algebra. In this case we give as an application the non-commutative translation, firstly studied in \cite{4}, that can be understood as an example of canonical transformation for the Snyder algebra.

\section{Non-commutative poisson brackets}

In general the Poisson brackets obey the following properties:

\begin{eqnarray}
\{ u, u \}  & = & 0 \nonumber \\
\{ u, v \}  & = & - \{ v, u \} \nonumber \\
\{ u, v w \}  & = & \{ u, v \} w + v \{ u, w \}
 \label{21}
\end{eqnarray}

In the case of one degree of freedom $( q,p )$ the canonical choice in phase space is introducing the fundamental Poisson brackets

\begin{eqnarray}
\{ q, q \}  & = & \{ p, p \} = 0 \nonumber \\
\{ q, p \}  & = &  1 \label{22}
\end{eqnarray}

from which the formula for computing the Poisson brackets for two generic phase space functions $u(q,p), v(q,p)$ is

\beq \{ u, v \} = \frac{\partial u}{\partial q} \frac{\partial v}{\partial p} \ - \ \frac{\partial u}{\partial p} \frac{\partial v}{\partial q}
\label{23}\eeq

Moreover the Jacobi identity holds

\beq \{ u, \{ v, w \} \} \ + \ \{ w, \{ u, v \} \} \ + \{ v, \{ w, u \} \} \ = 0 \label{24}\eeq

This property implies that if two constants of motion with respect to an Hamiltonian $ H(q,p)$ are given

\beq \dot{u} \ = \ \{ u, H \} \ = \ \dot{v} \ = \ \{ v, H \} \ = 0 \label{25}\eeq

then $ \{ u, v \} $ is also a constant of motion.

\beq \{ \{ u, v \}, H \} \ = \ 0 \label{26} \eeq

In this article we will discuss the properties of non-commutative Poisson brackets. In this case the choice of fundamental Poisson brackets is modified:

\begin{eqnarray}
\{ q, q \}  & = & \{ p, p \} = 0 \nonumber \\
\{ q, p \}  & = &  1 \ + \ \beta p^2  \label{27}
\end{eqnarray}

At a quantum level this choice is physically relevant because it produces an uncertainty in the position measurement

\beq \Delta q \ \sim \ \hbar^2 \beta  \label{28} \eeq

The non-canonical choice modifies the formula for computing the Poisson brackets of two generic functions $u(q,p), v(q,p)$

\beq \{ u, v \} \ = \ \left( \frac{\partial u}{\partial q} \frac{\partial v}{\partial p} \ - \ \frac{\partial u}{\partial p} \frac{\partial v}{\partial q}
\right) \ ( 1 \ + \ \beta p^2 ) \label{29}\eeq

keeping unchanged the other basic properties. The equations of motion maintain the same form

\beq \dot{q} \ = \ \{ q, H \} \  \ \ \ \ \  \ \dot{p} \ = \ \{ p, H \} \ = 0 \label{210}\eeq

We have noticed in ref \cite{5} that at lagrangian level it is possible to describe the same equations of motion, by changing the relation between Lagrangian and Hamiltonian

\beq {\rm \cal{L} } \ = \ - \ \frac{\dot{p} q }{  1 + \beta p^2 } \ - \ H(q,p) \label{211}\eeq

\section{$\beta$-canonical transformations}

In the formulation of quantum mechanics it is well known that the quantization rule

\beq [ \ \hat{q}, \hat{p} \ ] \ = \ i \hbar \ \{ q, p \} \ = \ i \hbar \label{31}
\eeq

is robust since it is stable with respect to some reparameterizations of phase space, which are called canonical transformations. These are defined by the requirement that the map

\beq ( q, p) \ \rightarrow \ ( \ Q(q,p,t), P(q,p,t) \ ) \label{32}\eeq

respects the fundamental Poisson brackets, i.e.

\beq \{ Q(q,p,t), P(q,p,t) \} \ = \ 1 \ = \ \frac{\partial Q}{\partial q} \frac{\partial P}{\partial p} \ - \ \frac{\partial Q}{\partial p} \frac{\partial P}{\partial q}
\label{33}\eeq

In general we have also the following property

\beq  \{ u, v \}_{(q, p)} \ = \ \{ u, v \}_{(Q, P)}
\label{34}\eeq

i.e. the calculation of a Poisson bracket is independent from the basic variables which parameterize the phase space, once that eq. (\ref{33}) holds.

Moreover the canonical transformations respect the Lie criterion, i.e.

\beq p \dot{q} \ - \ P \dot{Q} \ - \ H(q,p,t) \ + \ K(q,p,t) \ = \ \frac{d}{dt} F(q,p,t)
\label{35} \eeq

where the novel Hamiltonian $K(q,p,t)$ is defined as

\beq K \ = \ H \ + \ \frac{\partial F }{\partial t} \label{36}\eeq

The effect of the reparameterization on the Lagrangian is introducing a total derivative with respect to time that doesn't change the equations of motion.

The phase space measure

\beq dq \ dp \ \ \rightarrow \ \ dQ \ dP \label{37}\eeq

is invariant under canonical transformations.

Now if we want convince ourselves that the quantization rule must be modified with the introduction of a minimal fundamental length ( related to the non-commutative parameter $\beta$ ),
we must impose that this generalized quantization rule is stable with respect to a new class of phase space reparameterizations, i.e. imposing the invariance of the fundamental non-commutative Poisson brackets

\beq \{ q, p \} \ = \ 1 \ + \ \beta \ p^2  \ \ \rightarrow \ \ \{ Q, P \} \ = \ 1 \ + \ \beta \ P^2
\label{38}\eeq

from which we must obey the condition

\beq  \left( \frac{\partial Q}{\partial q} \frac{\partial P}{\partial p} \ - \ \frac{\partial Q}{\partial p} \frac{\partial P}{\partial q}
\right) \ ( 1 \ + \ \beta \ p^2 ) \ = \ 1 \ + \ \beta \ P^2
\label{39}\eeq

We will give in the following examples of such new class of reparameterizations, which we will call $\beta$-canonical transformations.

The other properties of the canonical transformations are modified as follows

i) the phase space measure invariant under $\beta$-canonical transformations is of the type

\beq \frac{dq \ dp}{ 1 \ + \ \beta \ p^2 } \ \ \rightarrow \ \frac{dQ \ dP}{ 1 \ + \ \beta \ P^2 } \label{310}\eeq

ii) the Lie criterion now reads

\beq -\frac{\dot{p} q}{ 1 \ + \ \beta \ p^2 } \ + \ \frac{\dot{P} Q }{ 1 \ + \ \beta \ P^2 } \ - \ H(q,p,t) \ + \ K(q,p,t) \ = \ \frac{d}{dt} F(q,p,t)
 \label{311}\eeq

the effect on the Lagrangian is identical to the standard case although we must take into account the different relation between Lagrangian and Hamiltonian

iii) at the quantum level the correct formula for the quantization of a generic $1d$ system invariant under $\beta$-canonical transformations is

\beq Z \ = \ \int \frac{dq \ dp}{ 1 \ + \ \beta \ p^2 } \ e^\frac{i S}{\hbar} \label{312}\eeq

where the action is defined as

\beq  S \ =  \ \int \ dt \ \left( -\frac{\dot{p} q}{ 1 \ + \ \beta \ p^2 } \  - \ H(q,p,t)
\right) \label{313}\eeq

With respect to the one derived in our previous paper \cite{5} we notice that the reparameterization invariance implies a different choice for the path integral measure.

\section{Harmonic oscillator with minimal length}

In the case of the standard harmonic oscillator, where $m$ is the mass and $\omega$ the pulsation, the solution in phase space can be parameterized in terms of the initial conditions

\begin{eqnarray}
q(t) & = & q(0) \ \cos \omega t \ + \ \frac{p(0)}{m \omega} \ sin \omega t
\nonumber \\
p(t) & = & p(0) \ \cos \omega t \ - \ m \omega \ q(0) \ sin \omega t \label{41} \end{eqnarray}

This transformation can be seen as a canonical transformation between the phase space variables

\beq ( \ q(0), p(0) \ ) \ \ \rightarrow \ \ ( \ q(t), p(t) \ )
\label{42}\eeq

Indeed

\beq \{ q(t), p(t) \} \ = \ \{ q(0), p(0) \} \ ( \ cos^2 \omega t \ + \ sin^2 \omega t \ ) \ = \ 1
\label{43}\eeq

We are going to show that in the case of the harmonic oscillator with a minimal length, the same phase space reparameterization (\ref{42}) can be interpreted as a $\beta$-canonical transformation, i.e. if

\beq \{ q(0), p(0) \} \ = \ 1 \ + \ \beta \ p^2(0)
\label{44}\eeq

then

\beq \{ q(t), p(t) \} \ = \ 1 \ + \ \beta \ p^2(t)
\label{45}\eeq

In this case the Hamiltonian maintains the same form

\beq H \ = \ \frac{p^2}{2m} \ + \ \frac{1}{2} m \omega^2 q^2  \ \ \ \ \ \ \{ q, p \} \ = \ 1 \ + \ \beta \ p^2  \label{46}\eeq

It gives rise to the following equations of motion

\begin{eqnarray}
\dot{q}(t) & = & \frac{p(t)}{m} \ ( 1 + \beta p^2 )
\nonumber \\
\dot{p}(t) & = & - m \omega^2 q(t) \ ( 1 + \beta p^2 )  \label{47} \end{eqnarray}

The solution of this system of equations has been presented in \cite{5} and can be written as

\begin{eqnarray}
q(t) & = &
\frac{\sqrt{ 2mE \  ( 1 + \beta \ 2 m E ) }}{ m \ \omega } \ \frac{ sin ( \omega t \sqrt{  1 + \beta \ 2 m E ) }}{ \sqrt{ 1 + \beta \ 2 m E \  sin^2 ( \omega t \sqrt{  1 + \beta \ 2 m E ) } } }
\nonumber \\
p(t) & = & \frac{ \sqrt{ 2mE } \ \cos ( \omega t \sqrt{  1 + \beta \ 2 m E ) }}{ \sqrt{ 1 + \beta \ 2 m E \  sin^2 ( \omega t \sqrt{  1 + \beta \ 2 m E ) } } } \label{48} \end{eqnarray}

Let us note that there are two ways to give the initial conditions, i.e. ($q(0), p(0)$) or ( $ E , t_0 $ ). In particular thanks to energy conservation the relation between these two types of initial conditions is

\begin{eqnarray}
E & = & \frac{p^2(0)}{ 2 m } \ + \ \frac{1}{2} \ m \omega^2 q^2(0)
\nonumber \\
t_0 & = & \frac{1}{\omega \ \sqrt{ 1 + \beta \ 2 m E }} \ \arctan \left[\frac{ m \omega}{\sqrt{  1 + \beta \ 2 m E }}  \ \frac{q(0)}{p(0)}
\right] \label{49} \end{eqnarray}

Let us compute the modified Poisson brackets of ($t_0, E$) in the basis ($q(0), p(0)$)

\begin{eqnarray}
\{ t_0, E \}_{(q(0), p(0))} & = & \left( \frac{\partial t_0}{\partial q(0)} \ \frac{\partial E}{\partial p(0)} \ - \ \frac{\partial t_0}{\partial p(0)} \ \frac{\partial E}{\partial q(0)}
\right) \ ( 1 + \beta p^2(0) ) =
\nonumber \\
 & = & \frac{1}{ 1 + \beta p^2(0) } \ ( 1 + \beta p^2(0) ) \ = \ 1 \label{410} \end{eqnarray}

therefore the pair of phase space variables ($t_0, E$) is canonical, even if the relation between $q$ and $p$ is not canonical. This is not surprising since we work in a non-relativistic approximation and time is not involved by the non-commutative deformation.

Now we recalculate the Poisson bracket between $q(t)$ and $p(t)$ in the basis of the canonical variables ( $t, E$ )

 \beq \{ q(t), p(t) \}_{(t, E)} \ = \ \left( \frac{\partial q(t)}{\partial t} \ \frac{\partial p(t)}{\partial E} \ - \ \frac{\partial q(t)}{\partial E} \ \frac{\partial p(t)}{\partial t}
\right) \label{411}\eeq

Let us substitute to $q(t)$

\beq q(t) \ = \ \frac{1}{ m \omega } \ \sqrt{ 2 m E \ - \ p^2(t) } \label{412}\eeq

from which

\beq \{ q(t), p(t) \}_{(t, E)} \ = \ - \ \frac{ \dot{p}(t) }{ m \omega^2 q(t) } \ = \ 1 + \beta p^2 (t) \label{413}\eeq

Therefore the temporal evolution is indeed a $\beta$-canonical transformation between the initial conditions ($q(0),p(0)$) and ($q(t),p(t)$).

\section{Generalization to higher degrees of freedom}

The generalization to higher degrees of freedom is obtained by substituting the algebra (\ref{27}) with the Snyder algebra

\begin{eqnarray}
\{ p_i, p_j \} & = & 0 \nonumber \\
\{ q_i, p_j \} & = & \delta_{ij} \ + \ \beta \ p_i p_j \nonumber \\
\{ q_i, q_j \} & = & \beta \ ( \  p_j q_i \ - \ p_i q_j \ )
 \label{51} \end{eqnarray}

This algebra can be generated by a canonical Poisson bracket through the following coordinate transformation

\begin{eqnarray}
q_i & = & \sqrt{1 - \beta w^2} \ z_i \ \ \ \ \ \ \ \ w^2 \ = w_i w_i \nonumber \\
p_i & = & \frac{w_i}{\sqrt{1 - \beta w^2}} \label{52} \end{eqnarray}

The Poisson bracket for two generic functions $u,v$ of the phase space generated by ($z_i, w_i$) can be written as

\beq \{ u, v \}_{( z_i, w_i )} \ = \ \sum^n_{i=1} \ \left(  \frac{\partial u}{\partial z_i} \  \frac{\partial v}{\partial w_i} \ - \
\frac{\partial v}{\partial z_i} \  \frac{\partial u}{\partial w_i} \right) \label{53}\eeq

In the canonical basis ($z_i, w_i$) the lagrangian clearly exists and it is defined as usual

\beq {\rm \cal{L}} \ = \ w_i \dot{z}_i \ - \ H( z_i, w_i ) \label{54} \eeq

By going to the basis ($q_i, w_i$), we notice that, similarly to the $1d$ case, the relation between Lagrangian and Hamiltonian is modified

\beq {\rm \cal{L}} \ = \ w_i \frac{d}{dt} \left( \frac{q_i}{\sqrt{1 - \beta w^2}} \right) \ - \ H(q_i, w_i) \ \sim \ - \frac{\dot{w}_i}{\sqrt{1 - \beta w^2}} q_i
 \ - \ H( q_i, w_i ) \label{55} \eeq

This Lagrangian must give rise to the same set of equations of motion of the Hamiltonian formalism with the non-commutative Poisson brackets. Now we must apply the coordinate transformation (\ref{52}) to determine the formula for the Poisson brackets $\{ u, v \}_{( q_i, p_i )}$ in the basis ($ q_i, p_i $). In general this gives rise to quite non-trivial formulas and we limit ourselves to the case of the intermediate basis ($ q_i, w_i $):

\begin{eqnarray}
q_i & = & \sqrt{1 - \beta w^2} \ z_i \ \ \ \ \ \ \ \ w^2 < \ \frac{1}{\beta} \nonumber \\
w_i & = & w_i  \label{56} \end{eqnarray}

As a consequence of this transformation the partial derivatives change in the following way

\begin{eqnarray}
\frac{\partial}{\partial z_i} & \rightarrow & \sqrt{1 - \beta w^2} \ \frac{\partial}{\partial q_i}  \nonumber \\
\frac{\partial}{\partial w_i} & \rightarrow & \frac{\partial}{\partial w_i} \ - \ \beta \ \frac{w_i q_j}{(1 - \beta w^2)} \ \frac{\partial}{\partial q_j} \label{57}
\end{eqnarray}

hence we obtain the final formula

\begin{eqnarray}
\{ u, v \}_{( q_i, w_i )}  & = & \sqrt{1 - \beta w^2} \ \sum^n_{i=1} \ \left(  \frac{\partial u}{\partial q_i} \  \frac{\partial v}{\partial w_i} \ - \
\frac{\partial v}{\partial q_i} \  \frac{\partial u}{\partial w_i} \right) \ +
 \nonumber \\
 & + & \beta \sum^n_{i,j=1} \ \left(  \frac{q_i w_j \ - \ q_j w_i}{\sqrt{1 - \beta w^2}} \ \right) \ \frac{\partial u}{\partial q_i} \
\frac{\partial v}{\partial q_j} \label{58} \end{eqnarray}

We notice that the non-commutative Poisson bracket is more complicated than the one dimensional case. We will need this formula to discuss a physically relevant case, i.e. the non-commutative translation of the Snyder algebra. In this intermediate basis we notice that a $\beta$-canonical transformation must respect the constraint $ w^2 < \ \frac{1}{\beta} $,
and therefore

\beq (q_i, w_i) \ \ \ w^2 < \ \frac{1}{\beta} \ \rightarrow \ (Q_i, W_i) \ \ \ W^2 < \ \frac{1}{\beta} \label{59} \eeq

\section{Non-commutative translation as a $\beta$-canonical transformation}

Obviously as an example of $\beta$-canonical transformations for several variables we could take the solution of a system of $N$ harmonic oscillators in a Snyder algebra. Instead we prefer
to give a more elementary example which is physically relevant, i.e. the case of non-commutative translation.

In the standard case simple translation of $q_i$ and $p_i$ are examples of canonical transformations.

Now we must satisfy the constraint that the fundamental Poisson brackets are independent from the basis choice

\begin{eqnarray}
\{ Q_i, W_j \}_{(q_i, w_i)} & = & \sqrt{1 - \beta W^2} \ \delta_{ij} \nonumber \\
\{ Q_i, Q_j \}_{(q_i, w_i)} & = & \beta \ \left(  \frac{Q_i W_j \ - \ Q_j W_i}{\sqrt{1 - \beta W^2}} \ \right)
\label{61} \end{eqnarray}

with the Poisson bracket written as in (\ref{58}) in the basis of the canonical variables ($q_i, w_i$). It is clear that a simple translation of $q_i$ cannot work, however we can satisfy the constraint (\ref{61}) in the following class of transformations

\begin{eqnarray}
Q_i & = & q_i \ + \ \xi_i \  f(w^2) \ + \ w_i \ ( w \cdot \xi ) \ g(w^2) \ \ \ \ \ \ \ \ f(w^2) \ + \ w^2 \ g(w^2) \ = \ 1 \nonumber \\
W_i & = & w_i \label{62} \end{eqnarray}

Without going into details, the constraint (\ref{61}) has already been solved in our previous article \cite{4}, from which we simply quote the result

\begin{eqnarray}
& \ & 2 f'(w^2) \ - \ g(w^2) \ + \ \frac{\beta}{1-\beta w^2} \ f(w^2) \ = \ 0 \nonumber \\
& \ & f(w^2) \ + \ w^2 \ g(w^2) \ = \ 1 \label{63} \end{eqnarray}

from which the solution is

\beq f(w^2) \ = \ \sqrt{\frac{1-\beta \ w^2}{\beta \ w^2}} \ \arcsin \sqrt{\beta \ w^2} \label{64} \eeq

The non-commutative translation of the Snyder algebra is indeed an example of $\beta$-canonical transformation. As a final check we compute

\beq - \frac{\dot{w}_i}{\sqrt{1 - \beta w^2}} q_i \ + \ \frac{\dot{W}_i}{\sqrt{1 - \beta W^2}} Q_i \ = \ \frac{d}{dt} \ \left( \frac{ ( \xi \cdot w ) \ \arcsin \sqrt{\beta \ w^2} }{\sqrt{\beta \ w^2}} \right) \label{65} \eeq

that satisfies the Lie criterion, and the Lagrangian equations of motion are not affected by the non-commutative translation.

\section{Conclusions}

In this article we have clarified the role of non-commutativity at a classical level. After discussing the properties of non-commutative Poisson brackets, we have introduced the innovative concept of $\beta$-canonical transformation, as phase space reparameterization that maintains the definition of fundamental Poisson brackets. The same definition of the non-commutative path integral with one degree of freedom respects exactly this invariance and probably the same property holds with higher degrees of freedom after introducing the Snyder algebra.

We have given practical examples of such new class of transformations like the temporal evolution of the $1d$ harmonic oscillator with minimal length and the non-commutative translation of the Snyder algebra.

We believe that the development of this new type of invariance will be useful for studying the quantum systems with minimal length. Our ultimate aim would be to have full control of the non-commutative version of quantum field theory.

\end{document}